\documentclass{WileyMSP-template}

\usepackage[sorting=none]{biblatex}
\usepackage{textcomp}
\begin{document}

\pagestyle{fancy}
\setlength{\headheight}{25pt}
\rhead{\includegraphics[width=2.5cm]{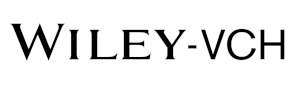}}

\title{Probing near-field EM fluctuations in superparamagnetic CoFeB with NV quantum dephasometry}

\maketitle


\author{Shoaib Mahmud$^{1,2,*}$,}
\author{Wei Zhang$^{1,2,*}$,}
\author{Pronoy Das$^{1,2}$,}
\author{Angshuman Deka$^{2}$,}
\author{Wenbo Sun$^{1,2}$,}
\author{Zubin Jacob$^{1,2}$}


\dedication{*Equal Contribution}


\begin{affiliations}
$^1$Elmore Family School of Electrical and Computer Engineering, Purdue University, West Lafayette, Indiana 47907, USA.\\
$^2$Birck Nanotechnology Center, and Purdue Quantum Science and Engineering Institute, West Lafayette, Indiana 47907, USA.\\

\end{affiliations}



\begin{abstract}

\justifying
Superparamagnetism in nanoscale magnetic layers is a critical property for a wide range of spintronic-based sensor and computing applications. While conventional magnetization measurements can detect superparamagnetic signatures, they often require the application of high perturbative fields and are difficult to implement for magnetic layers integrated within functional devices. In this study, we non-invasively investigate the superparamagnetic spin dynamics of a nanoscale CoFeB layer of thickness 1.1 nm, deposited on a diamond substrate, by probing its low-frequency near-field electromagnetic (EM) fluctuations using nitrogen-vacancy (NV) centers-based quantum dephasometry. Our measurements reveal an unconventional, non-monotonic temperature dependence of the dephasing time of NV centers, which we attribute to EM fluctuations produced by thermally driven superparamagnetic domain flipping in CoFeB. Our findings are further supported by the theoretical interpretation of the dephasing dynamics of NV centers and the complementary SQUID-based magnetization characterizations of the CoFeB layer. Additionally, exploiting the technique of NV dephasometry, we extract the spectral density of the EM fluctuations in CoFeB, which is used to isolate different components of the EM fluctuations acting on NV centers. We also measure the CoFeB-to-NV distance-dependent coherence times of NV centers to investigate the effect of the dimensionality of the CoFeB layer on the generated near-field EM fluctuations. These results provide critical insight into the magnetization dynamics and near-field EM environment of nanoscale magnetic layers. It also has significant implications for the development of hybrid quantum spintronic devices and applications involving nanoscale opto-magnetic materials.

\end{abstract}

\section{Introduction}

\justifying
Magnetic thin layers are widely used in various applications, including sensing \cite{ikeda2010perpendicular, dong2022thin, bauer2025exploiting, feng2010superparamagnetism}, data storage \cite{tudu2017recent, bhatti2017spintronics}, and spintronic devices \cite{xu2025thin, wang2017spintronic}. CoFeB is a commonly employed magnetic material due to its high tunneling magnetoresistance (TMR) when used as magnetic tunnel junctions (MTJs), its highly tunable magnetic anisotropy, and high saturation magnetization \cite{ikeda2010perpendicular,liu2025linear, nakano2016magnetic}. Ultrathin-layered CoFeB has found practical applications in spintronic devices such as, Poisson bolometer \cite{bauer2025exploiting, mousa2026ultra, singh2025long}. At the nanoscale, its magnetic properties can differ significantly from those of the bulk counterpart, especially when the size is comparable to that of magnetic domains. One such unconventional property that arises due to reduced dimensionality is superparamagnetism \cite{voogt1998superparamagnetic, bedanta2008supermagnetism, jang2006magnetic}, where the magnetization of the material can randomly flip due to thermal fluctuations. This phenomenon is useful in sensing and spintronics applications, such as bolometers and biomarker detection \cite{bauer2025exploiting,masud2019superparamagnetic,  neuberger2005superparamagnetic}. This is also important in determining the limiting factors in magnetic storage devices and density-material design \cite{vopson2014multiferroic, mohapatra2023low, weller2000extremely}. Superparamagnetism can strongly depend on the deposition and synthesis methods of the material, as well as the type of device in which the material is integrated. So, to probe superparamagnetic behavior in magnetic materials reliably, a sensitive probing methodology is needed after material integration into the device. 

NV center serves as an excellent quantum sensor for the characterization of the magnetic properties of materials, including ferromagnets \cite{lee2021quantum, solanki2022electric}, antiferromagnets \cite{wang2022noninvasive}, and van der Waals magnets \cite{huang2023layer, huang2023revealing, huang2022wide}. NV center-based quantum noise spectroscopy has proven to be effective in measuring critical fluctuations in different materials \cite{machado2023quantum, li2024critical}. This is attributed to the versatility of NV center, which can operate over a wide temperature range -  from cryogenic to above room temperature, can be placed in close proximity to the system of interest, and offer broad dynamic sensitivity, ranging from DC to GHz frequencies. Spectroscopy based on NV relaxometry \cite{lee2021quantum, mclaughlin2022quantum, kumar2024room}, which probes EM fluctuations in the frequency range close to the NV resonant frequency ($\approx$ 2.87 GHz), has successfully captured numerous phenomena. However, recent studies have highlighted the limitations of this technique in detecting low-frequency MHz-range physics. Specifically, phenomena such as complex vortex-solid fluctuations \cite{liu2025quantum}, and magnon hydrodynamics in two-dimensional materials  \cite{xue2024signatures} are not well resolved by conventional relaxometry techniques. The Quantum dephasometry technique of probing MHz EM fluctuations is quite useful in capturing these low-frequency phenomena. 

In this work, we study the near-field low-frequency EM fluctuations of nanoscale CoFeB layers using NV quantum dephasometry. We demonstrate that a 1.1 nm thick CoFeB layer generates MHz-range EM fluctuations caused by superparamagnetic domain flipping, which we capture in the dephasing rate of NV centers. We also provide a theoretical framework to interpret the observed effect of superparamagnetic fluctuations on the dephasing time of NV centers. Complementary SQUID-based magnetization characterizations are done for the nanoscale CoFeB layer. Using quantum dephasometry, we measure the low-frequency spectral density of the EM fluctuations in CoFeB. Finally, we explore the distance dependence of the superparamagnetic near-field fluctuations on the NV dephasing time. Our methodology not only provides a path to probe exotic magnetic phases in integrated devices but also paves the way for the development of hybrid quantum spintronic devices.  This approach offers significant advantages and can be readily applied to noise spectroscopy in a wide range of spintronic devices.

\section{Results and Discussions}
\subsection{Superparamagnetism in nanoscale CoFeB layer}

\begin{figure}[ht!]
\centering\includegraphics[scale = 0.62]{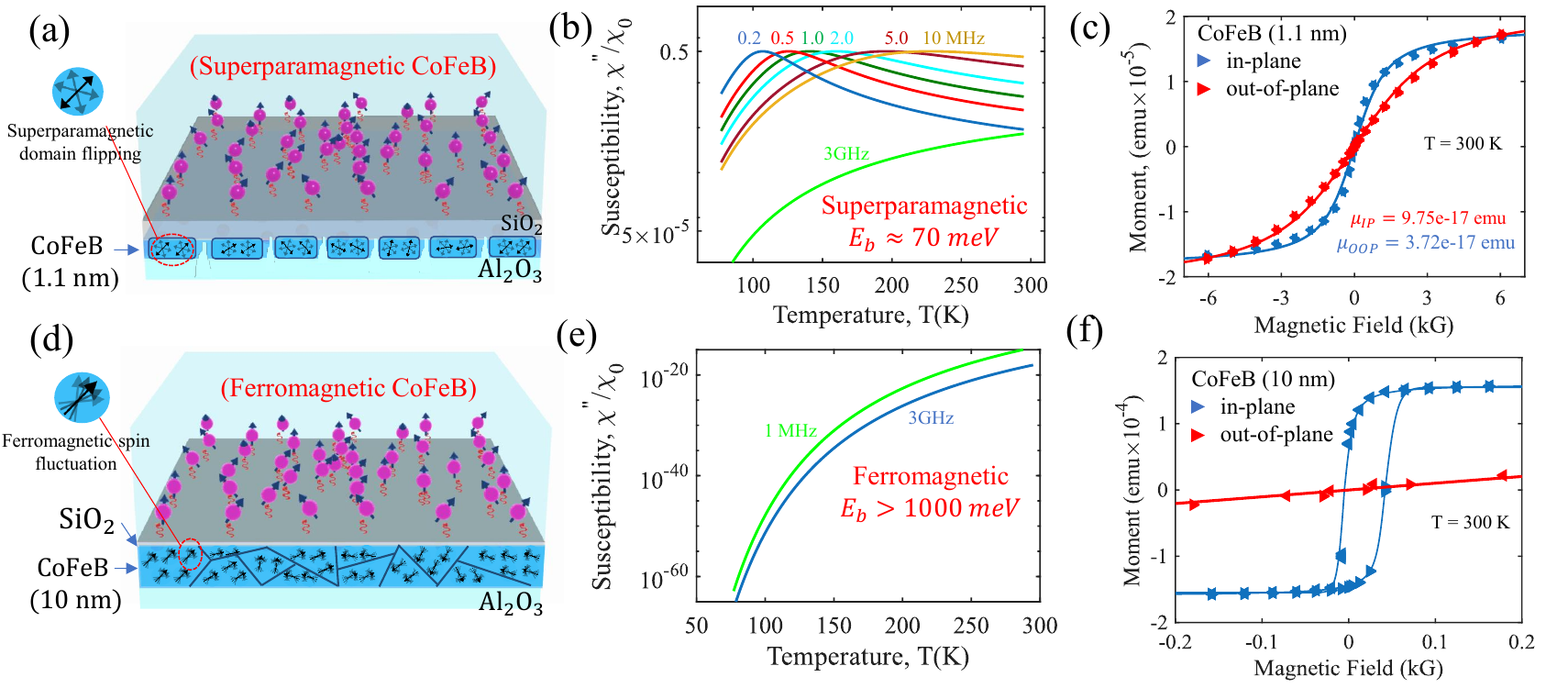}
\caption{Magnetization responses of nanoscale CoFeB layers. (a) A 1.1 nm thick CoFeB deposited diamond sample with an embedded NV layer. For a CoFeB layer with such thickness, superparamagnetism arises due to the formation of nanoscale magnetic elements with a smaller anisotropy barrier energy ($E_b$). (b) Frequency dependence of the magnetic susceptibility in superparamagnetic CoFeB. Non-monotonicity appears in MHz frequency response due to the transition from the superparamagnetic state to the blocked state. The temperature corresponding to the susceptibility peak value is termed as the blocking temperature. The transition point for GHz susceptibility is beyond the plotted temperature range. (c) SQUID-based magnetization curves of superparamagnetic CoFeB. $\mu_{IP}$ and $\mu_{OOP}$ are the average moments of the nanoscale magnetic elements along the in-plane and out-of-plane directions, respectively. (d) A diamond sample with a 10 nm thick CoFeB layer. Ferromagnetism is retained in this continuous magnetic layer due to a higher anisotropy barrier energy. (e) MHz and GHz frequency responses of magnetic susceptibility for ferromagnetic CoFeB. For both cases, there is no phase transition in the magnetization state. (f) SQUID-based magnetization curve for the ferromagnetic CoFeB layer.}
  \label{figure1N}
\end{figure}
Superparamagnetism is a phenomenon that occurs in nanoscale magnetic elements, where magnetization can be randomly flipped due to thermal effects. In a thin CoFeB layer (thickness $<$ 1.5 nm), discontinuities within the layer can give rise to nanoscale magnetic elements \cite{feng2010superparamagnetism}. The structure of our device with a nanoscale CoFeB layer (thickness $\approx$ 1.1 nm) is shown in Fig. \ref{figure1N}(a). Superparamagnetism arises due to the reduced anisotropy energy barrier ($E_b$) for domain flipping in the nanoscale magnetic elements. This superparamagnetic behavior can be quantitatively modeled using the following equation, which describes the thermally activated magnetic domain flipping rate as a function of temperature \cite{kaneko2024temperature}-
\begin{equation}
    f_N = f_0\ exp(-E_b/k_bT).
    \label{Neel_Arrh}
\end{equation}
Here, $f_0$ is the attempt frequency (typically 100 MHz – 10 GHz for CoFeB) \cite{kanai2021theory}, $E_b$ is the energy barrier for domain reversal (typically in the range of 20–200 meV), and $k_b$ is the Boltzmann constant. The temperature dependence of the frequency response of magnetic susceptibility in a superparamagnetic material is governed by domain dynamics and is given by \cite{singh2009ac} - 
\begin{equation}
Im[\chi(\omega)] = \chi_0 \frac{\omega \tau}{1 + (\omega \tau)^2},
\label{chi2_freq}
\end{equation}
where, $\chi_0$ is the static susceptibility in the low-frequency limit, and $\tau = 1/(2f_N)$ is the characteristic domain relaxation time at different temperatures. Fig. \ref{figure1N}(b) shows the temperature-dependent frequency response of magnetic susceptibility. The energy barrier $E_b$ is estimated to be 70 meV from theoretical modeling of the NV dephasing rate as discussed in section 2.3. As shown in the graph, the sub-10 MHz frequency response of the susceptibility exhibits a pronounced peak. The temperature corresponding to this maximum is known as the blocking temperature $T_B$. This marks the transition from thermally activated superparamagnetic behavior to a magnetically blocked state \cite{andersson1997monte}. In magnetically blocked state, the magnetization relaxes slowly due to reduced thermal activation effect. In the superparamagnetic state, the magnetization of CoFeB shows an S-shaped reversible curve, as shown in the SQUID measurements in Fig. \ref{figure1N}(c). These curves can be well-fitted using the Langevin function \cite{henrard2019monitoring}, which serves as a complementary confirmation for the superparamagnetic nature of our CoFeB layer.

When the thickness of the CoFeB layer is relatively higher, such as 10 nm, then the entire layer acts as a continuous ferromagnetic film, as shown in Fig. \ref{figure1N}(d). In the ferromagnetic regime, due to the higher anisotropy energy barrier, the susceptibility arising from magnetization flipping at MHz and GHz frequencies is much lower around room temperature, as shown in Fig. \ref{figure1N}(e).   Fig. \ref{figure1N}(f) shows the magnetization curve for ferromagnetic CoFeB. The smaller saturation magnetization field and the presence of a hysteresis loop in the in-plane direction arise due to ferromagnetic behavior in the thicker CoFeB film. Comparing the frequency dependence of the magnetic susceptibility (Figs. \ref{figure1N}(b) and \ref{figure1N}(e)), it is clear that superparamagnetism and ferromagnetism are fundamentally different phenomena. Through our measurements, we show that it is possible to get a quantitative idea about the frequency responses of superparamagnetic and ferromagnetic materials using NV quantum spectroscopy.

\subsection{NV-based quantum relaxometry vs. quantum dephasometry}

\begin{figure}[ht!]
\centering\includegraphics[scale = 0.58]{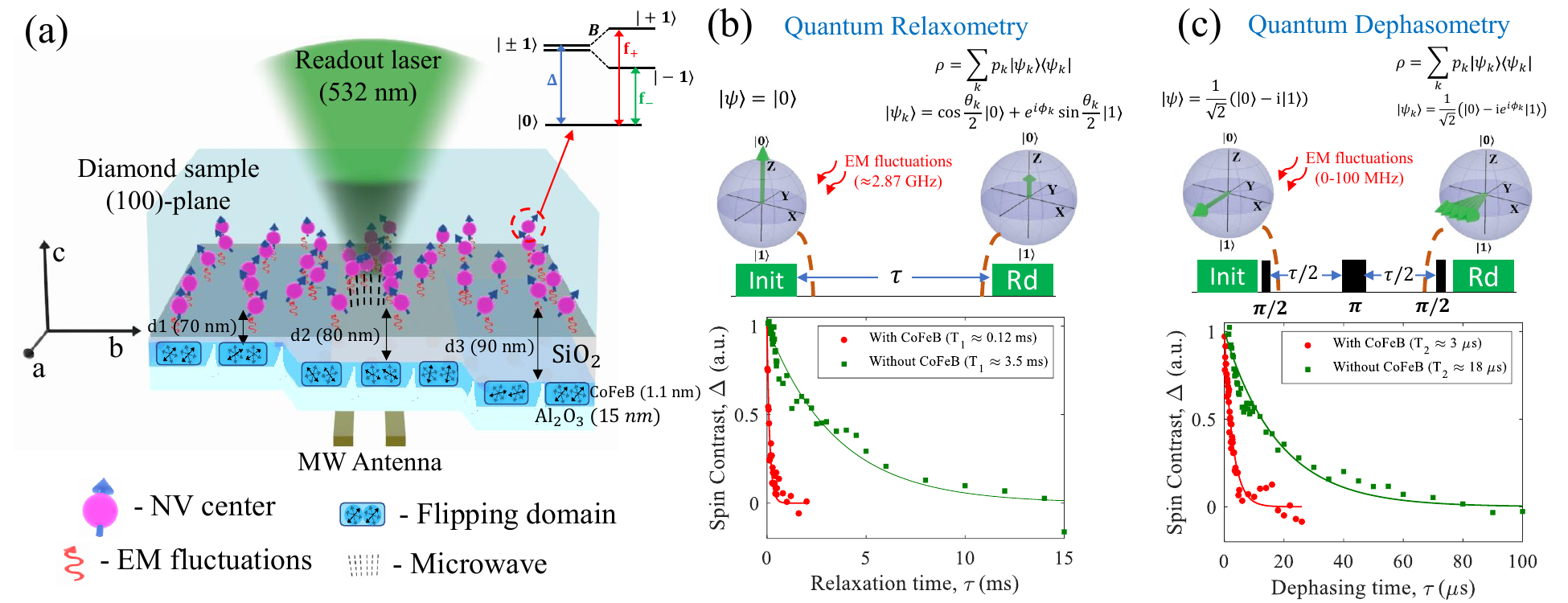}
\caption{NV-based quantum spectroscopy of near-field EM fluctuations: Relaxometry vs. Dephasometry. (a) Schematic illustration of the spectroscopy of near-field EM fluctuations with NV center spin qubits in diamond. A CoFeB layer is deposited on the diamond sample, with variable distances (d1, d2, d3) from the NV layer. A 532 nm green laser is used for optical initialization and readout of the NV spin states. An MW antenna delivers control signals for spin manipulation during dephasometry measurement. The ground state spin energy levels of NV center are shown at the top. (b) Pulse sequence and spin state evolution in the quantum relaxometry protocol. The spin is initialized into a pure state along the quantization axis, which undergoes relaxation due to interactions with EM fluctuations components resonant with the NV transition frequency ($\approx 2.87\ GHz$). The relaxation time $T_1$ is extracted from the exponential decay of spin polarization. The presence of proximal CoFeB enhances the relaxation rate, as shown in the accompanying data. (c) Schematic of the quantum dephasometry protocol. Here, the NV spin is initialized into a superposition state oriented perpendicular to the quantization axis. The phase coherence of the superposition state is sensitive to low-frequency, non-resonant EM fluctuations. The dephasing time $T_2$ is extracted from the decay rate of the phase coherence.}
  \label{figure1}
\end{figure}

Our approach to the spectroscopy of near-field EM fluctuations using NV center spin qubits is schematically illustrated in Fig. \ref{figure1}(a). A nanoscale CoFeB layer (thickness $\approx$ 1.1 nm; composition: Co-60\%, Fe-20\%, B-20\%) is deposited on a (100)-oriented diamond substrate. A layer of NV centers is created via ion implantation (ion concentration $\approx 10^{12}$ cm$^{-2}$) at an approximate depth of 50 nm from the surface. A SiO$_2$ spacer layer is used to control the separation between the CoFeB and the NV layer. The near-field EM fluctuations originating from the CoFeB layer influence the coherence properties of the NV spin states. A 532 nm green laser is used for optically reading out the NV spin states, while a microwave (MW) antenna provides the signal for coherent manipulation of NV centers.  

The NV center is a point defect in diamond with a spin-1 configuration, where the degenerate $\vert\pm1\rangle$ states are separated from the $\vert0\rangle$ state by a zero-field splitting of approximately 2.87 GHz at room temperature. The NV spin Hamiltonian in the presence of an external magnetic field is given by \cite{barry2020sensitivity} -
\begin{equation}
    H = \hbar\Delta {S_z}^2+\hbar\gamma\vec{B}.\vec{S},
    \label{NV_Hamilton}
\end{equation}
where, $\Delta\approx$ 2.87 GHz is the zero-field splitting, $\gamma$ = 28 GHz/T is the gyromagnetic ratio, $\hbar$ is the reduced Planck's constant, $\vec{B}$ is the external magnetic field, and $\vec{S}=(S_x,S_y,S_z)$ is the spin-1 operator. Taking the NV quantization axis along the z-direction, the effect of the external magnetic field on the spin Hamiltonian can be decomposed into longitudinal ($\hbar\gamma B_z S_z$) and transverse ($\hbar\gamma (B_x S_x+B_y S_y)$) components. The resonant energy-level splitting $f_\pm$ between the $\vert0\rangle$ and $\vert\pm1\rangle$ spin states (Fig. \ref{figure1}(a)) in the presence of a magnetic field $\vec{B}$ can be obtained by solving equation \ref{NV_Hamilton}.

The dynamics of NV spin states in the presence of a fluctuating magnetic field $\delta\vec{B}(t)$ depend on the frequency and directional components of the field. The qubit decoherence profile $e^{-f(t)}$ in the presence of noise can be written as \cite{bar2012suppression}- 
\begin{equation}
    f(t)=\int^\infty_0 \frac{d\omega}{\pi}F_\tau(\omega)S(\omega),
    \label{Spin_Decoher}
\end{equation}
where, $F_\tau(\omega)$ is the noise filter function that determines frequency selectivity, and $S(\omega)$ is the spectral density of the fluctuating field $\delta\vec{B}(t)$. 
When NV center spin qubit is polarized in a pure state ($\vert0\rangle$ or $\vert\pm1\rangle$), any transverse field fluctuation $\delta \vec{B}_{xy}(t)$ that is resonant with the NV spin transition frequency ($F_\tau(\omega)\approx\delta(\omega-2\pi f_\pm)$), induces spin-state relaxation, as depicted in Fig. \ref{figure1}(b). This relaxation results from resonant energy transfer between the fluctuating field and the NV center. The NV center decays into a mixed state $\rho$ during this process. The characteristic decay rate is quantified by $T_1$ time, also known as relaxation time, which corresponds to the time at which the NV state decays to $1/e$ of its initial population. The fluctuating field in a nanophotonic environment can be modeled using Green's function as, $\langle\delta B^2\rangle=\frac{\omega^2}{c^2}\mathrm{coth}(\frac{\hbar\omega}{2k_B T})\mathrm{Im}[\mathbf{G_m}(\mathbf{r},\omega)+\mathbf{G^\dagger_m}(\mathbf{r},\omega)]$\cite{premakumar2017evanescent}. Thus, the $T_1$ decay rate can be expressed as \cite{cortes2017super, novotny2012principles} - 
\begin{equation}
    \frac{1}{T_1} = \sum_{\omega_\pm} \frac{2\mu_o\omega^2}{\hbar c^2}\mathrm{coth}\left(\frac{\hbar\omega}{2k_B T}\right) \mathbf{m}.\mathrm{Im}[\mathbf{G_m}(\mathbf{r},\omega)+\mathbf{G^\dagger_m}(\mathbf{r},\omega)].\mathbf{m}
    \label{T1_Green}
\end{equation}
where, $\mu_0$ is the vacuum permeability and $\mathbf{m}$ is the dipole orientation vector. In a diamond sample, $T_1$ is primarily limited by intrinsic phonon interactions and spin-bath-induced noise. However, the presence of external, material-induced EM fluctuations can further reduce $T_1$, enhancing the NV decay rate - as shown in the graph of Fig. \ref{figure1}(b). Therefore, $T_1$ serves as a quantitative measure of near-field resonant EM fluctuations, which is the basis of quantum relaxometry. While this technique has been widely applied to probe material properties and phase transitions, it is insensitive to low-frequency EM fields. Many of these are associated with important phenomena such as superparamagnetism, vortex dynamics, and magnon hydrodynamics.

To access such low-frequency EM fluctuations ($F_\tau(\omega<<2\pi f_\pm)$), we employ quantum dephasometry, as shown in Fig. \ref{figure1}(c). In this method, the spin qubit is initialized into a coherent superposition state $\vert\alpha\rangle=(1/\sqrt{2})(\vert 0 \rangle-i\vert 1 \rangle)$ using a microwave $\pi/2$-pulse. In this state, the qubit becomes sensitive to any perturbative low-frequency longitudinal field fluctuations $\delta\vec{B}_z(t)$. The $\pi$ pulse is applied during the wait time $\tau$ to modulate the effect of the fluctuating field on the dephasing rate. In this protocol, no energy transfer occurs between the EM field and the NV center. As a result, the NV remains in a superposition state during $\tau$, but its phase $\phi$ accumulates noise-induced fluctuations. The resulting phase decay is read out from the projection of the final spin state $\rho$. The inverse of this decay rate defines the $T_2$ time, also known as the dephasing time. The dephasing rate can be expressed in terms of the Green's function at the material interface \cite{sun2025nanophotonic}- 
\begin{equation}
    \frac{1}{T_2} = \int_0^{\omega_\tau} \frac{2\mu_o\omega^2}{\hbar c^2} \mathrm{coth}\left(\frac{\hbar\omega}{2k_B T}\right)\mathbf{m}.\mathrm{Im}[\mathbf{G_m}(\mathbf{r},\omega)+\mathbf{G^\dagger_m}(\mathbf{r},\omega)].\mathbf{m}\ F_\tau(\omega)d\omega.
    \label{T2_Green}
\end{equation}
In diamond, NV dephasing is mainly caused by dipolar-coupled magnetic noise from nuclear spins \cite{bauch2020decoherence, park2022decoherence} and divacancy centers \cite{favaro2017tailoring}. However, in proximity to materials with significant low-frequency noise, the dephasing rate can increase, as shown in the graph of Fig. \ref{figure1}(c). Thus, $T_2$ provides a quantitative measure of near-field low-frequency EM fluctuations, which we utilize to probe the superparamagnetism of CoFeB in our measurements.

\subsection{Temperature dependence of the superparamagnetic EM fluctuations in CoFeB}
\begin{figure}[!ht]
\centering\includegraphics[scale = 0.65]{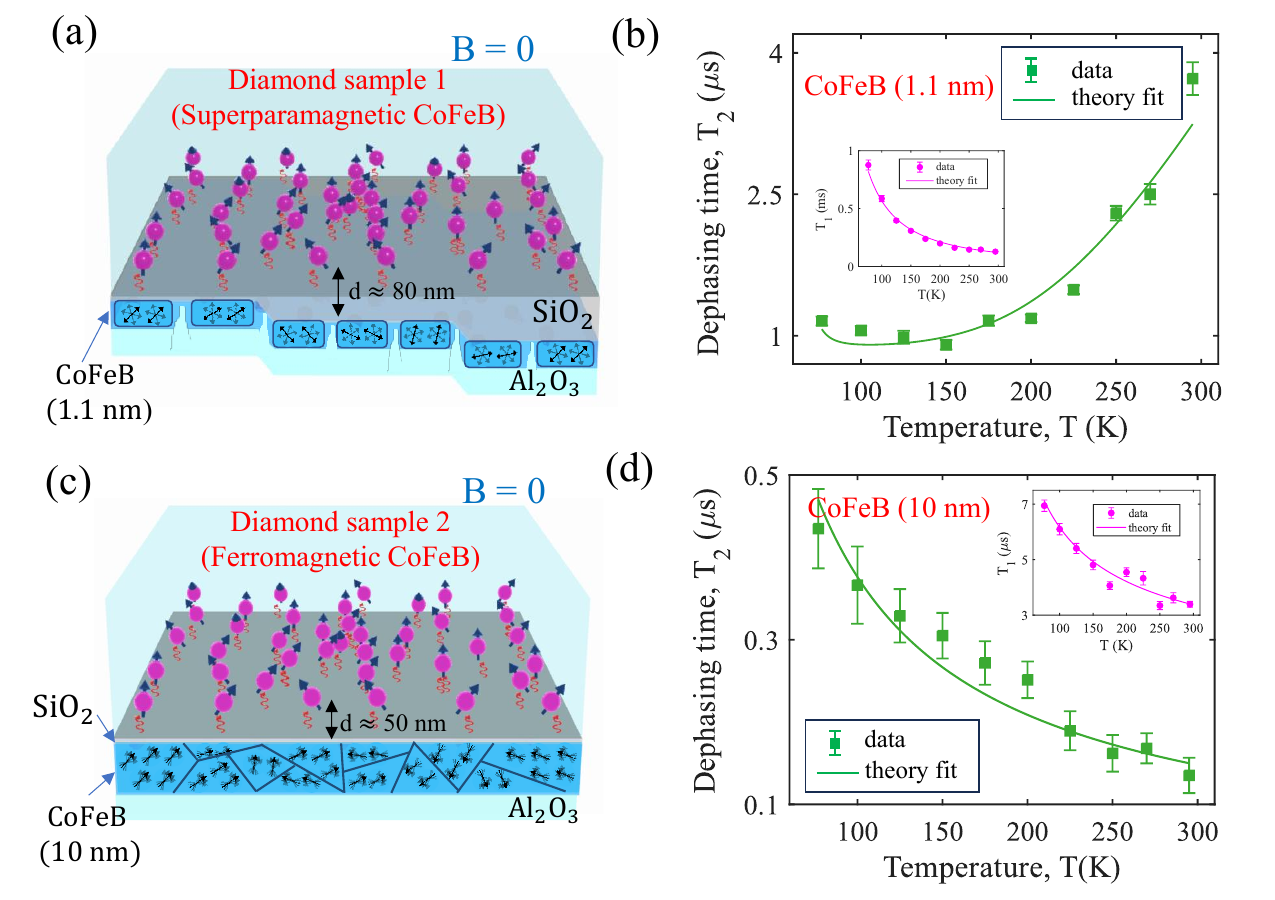}
\caption{Probing CoFeB superparamagnetism via temperature-dependent coherence measurements of NV centers. (a) Schematic illustration of the diamond sample with an ultrathin CoFeB layer showing superparamagnetic behavior. (b) Dephasing measurements of NV centers near the CoFeB layer. The temperature-dependent scaling behavior of this low-frequency noise-sensitive measurement highlights the differences with that of relaxation measurements shown in the inset. (c) Diamond sample with a thicker CoFeB layer showing ferromagnetic behavior. (d) Dephasing measurement near the ferromagnetic CoFeB layer. The relaxation measurement in shown in the inset.}
  \label{figure2}
\end{figure}

We first perform a temperature-dependent study of the coherence times of the NV centers, since superparamagnetism is a thermally driven phenomenon. The measurements are conducted at a distance of 80 nm from the superparamagnetic CoFeB layer, as shown in Fig. \ref{figure2}(a). The temperature dependence of the dephasing time $T_2$ of the NV centers is shown in Fig. \ref{figure2}(b). We observe a nonmonotonic behavior in this temperature dependence of the dephasing time. It decreases with temperature, reaching a minimum around 150 K, and starts to increase. This is unconventional as thermal-driven fluctuations are expected to reduce with temperature, which will increase the coherence time. This type of coherence increase is observable for the relaxation time $T_1$ shown in the inset. This unconventional scaling of the $T_2$ time is an indication of a critical phenomenon associated with a magnetic phase transition. To understand this temperature dependence, we examine the role of the EM fluctuations in determining the coherence times of the NV centers. The spectral density of the EM fluctuations, $S(\omega)$, is related to the dissipative part of the magnetic susceptibility, $\chi(q_{NV},\omega)$ via the fluctuation-dissipation theorem \cite{casola2018probing}-

\begin{equation}
S(\omega)\propto\frac{T}{\omega}\ Im[\ \chi(q_{NV},\omega)].
\label{Noise_Chi2}
\end{equation}
Here, $q_{NV} = 1/d_{NV}$, where $d_{NV}$ is the distance between the layer of NV centers and CoFeB. Substituting this into equation \ref{Spin_Decoher}, we obtain - 

\begin{equation}
\frac{1}{T_{1,2}} \propto \int_0^\infty \frac{T}{\omega} \,Im[\chi(q_{NV}, \omega)] \, F_\tau(\omega) \, d\omega.
\label{T12_Chi2}
\end{equation}
Thus, both relaxation ($T_1$) and dephasing ($T_2$) times depend on the frequency-dependent magnetic susceptibility and temperature. However, as discussed in section 2.2, these coherence times are sensitive to different components of the noise spectrum constrained by the noise filter function, $F_\tau(\omega)$.

Dephasing time $T_2$ is dependent on the MHz frequency responses of the magnetic susceptibility. From the magnetic susceptibility curves of superparamagnetic CoFeB shown in Fig. \ref{figure1N}(b), we observe a phase transition from superparamagnetic to blocked states around 100 - 160 K for MHz range responses. Our observed non-monotonic change in the dephasing time, $T_2$, can be explained by the trend of the change of magnetic susceptibility. The theoretical fitting shown in Fig. \ref{figure2}(b) is done using equation 8 and considering the magnetic susceptibility $Im[\chi]$ from Fig. \ref{figure1N}(b), and the filter function $F_\tau(\omega)$ corresponding to the pulse sequence in Fig. \ref{figure1}(c) [shown in supplementary document]. For EM fluctuations in the GHz range, which is much faster than the characteristic domain flipping rates of the CoFeB film, the susceptibility has a weak temperature dependence \cite{gittleman1974superparamagnetism}. Therefore, the relaxation time $T_1$ exhibits an approximately inverse temperature dependence.

This type of non-monotonic behavior is not observable in the ferromagnetic thick film case (sample 2 in Fig. \ref{figure2}(c)), where both the relaxation time $T_1$ (Fig. \ref{figure2}(d)) and dephasing time $T_2$ (shown in figure inset), follow a decreasing trend with increasing temperature. This is expected for the decay times of a spin qubit in the vicinity of an ordered magnetic material. All these experimental results and theoretical modeling demonstrate that our measurements effectively capture the frequency responses of different phases of magnetic materials.

\subsection{Low-frequency spectral analysis of EM fluctuations in CoFeB}

We perform a spectral analysis of the near-field EM fluctuations induced by the CoFeB layer to understand the dynamics and roles of its different components. For this analysis, we use an engineered dynamical decoupling pulse sequence \cite{bylander2011noise} [see the supplementary document for more information]. Fig. \ref{figure3}(a) presents the spectra of the EM fluctuations measured by NV centers in diamonds without and with CoFeB deposition. To explain the sources of the fluctuations, we perform a theoretical analysis of the spectrum. We can categorize the sources of EM fluctuations into 3 types - 1. Fluctuations due to intrinsic nuclear spin and divacancy centers $S_{intr}(f)$, 2. Fluctuations caused by superparamagnetic domain flipping $S_{dom}(f)$, 3. An uncorrelated source of fluctuations due to other spin noises and phonons $S_{other}(f)$. The total noise thus can be written as -
\begin{equation}
S(f) = S_{intr}(f)+S_{dom}(f)+S_{other}(f).
\label{Noise_All}
\end{equation}

EM fluctuations due to nuclear spins and divacancy centers can be modeled by a Lorentzian spectral density of the form \cite{bar2012suppression} -
\begin{equation}
S_{intr}(f) = \frac{D^2 \tau_c}{\pi} \frac{1}{1 + (2\pi f \tau_c)^2},
\label{lorentzFit}
\end{equation}
where, $D$ is the coupling strength of EM fluctuations to NV center spin qubits and $\tau_c$ is the correlation time of the fluctuations.

\begin{figure}[h!]
\centering\includegraphics[scale = 0.55]{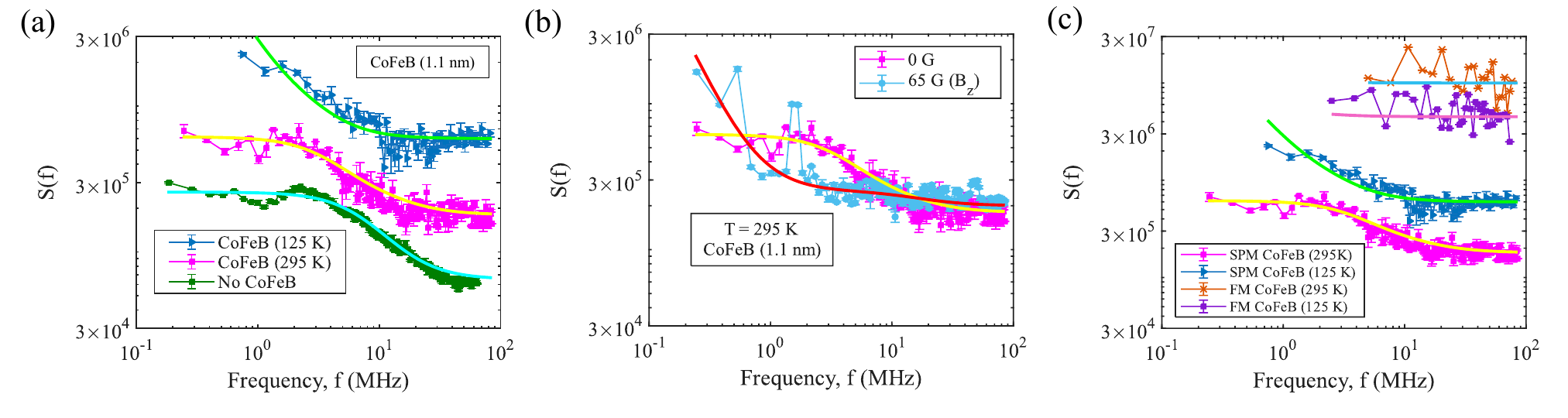}
\caption{Low-frequency spectroscopic characterization of the EM fluctuations from CoFeB film. (a) Spectra near CoFeB film at different temperatures. The intrinsic EM fluctuations spectrum (without CoFeB) at 295 K is primarily dominated by the nuclear spin bath and divacancy centers formed during implantation. In contrast, the presence of the CoFeB film induces added fluctuations. The theoretical fittings are plotted on the experimental data using a continuous bold line. (b) Effect of magnetic field on the CoFeB-induced fluctuations at 295 K. The applied magnetic field modifies the sub-10 MHz fluctuation component, likely due to changes in the diamond nuclear spin bath. However, the fluctuations above 10 MHz remain largely unaffected, consistent with the weak magnetic-field dependence of the EM fluctuations in superparamagnetic CoFeB. (c) Comparison of the spectra of superparamagnetic and ferromagnetic CoFeB layers at different temperatures. }
  \label{figure3}
\end{figure}

Fluctuations due to the superparamagnetic domain flipping $S_{dom}(f)$ can be modeled by inserting the frequency dependence of the susceptibility (equation \ref{chi2_freq}) into equation \ref{Noise_Chi2}. It can be written as,
\begin{equation}
S_{dom}(f) = \frac{\Delta_{SP}^2\tau_d}{1 + (2\pi f \tau_d)^2},
\label{Fluc_Fit_SP}
\end{equation}
where, $\Delta_{SP}$ describes the interaction strength of fluctuations with the NV centers, $\tau_d$ is related to the superparamagnetic domain flipping rate. The other uncorrelated sources of fluctuations $S_{other}(f)$ are considered to be frequency independent.

\begin{table}[h!]
    \centering
    \begin{tabular}{c|c|c|c|c|c|c|c}
        Sample &  T(K) & $B_z$(G) & D (MHz) & $\tau_c$(ns) & $\Delta_{SP}$(MHz) & $\tau_d$(ns) & $S_{other}$\\
        \hline \hline
        No CoFeB  & 295 & 0 & 5.5 & 20  & - & - & 6.5e4 \\ \hline
          & 295 & 0 & 5.5 & 40  & 4   & 10  & 1.8e5 \\
        Superparamagnetic CoFeB  & 295 & 65 & 5.5 & 2000  &  4 & 10  & 2e5 \\
          & 125 & 0 & 5.5 & 60  &  11 &  500 & 6e5 \\
        \hline   
        Ferromagnetic CoFeB  & 295 & 0 & 5.5 & 100  & - & - & 1e7 \\
          & 125 & 0 & 5.5 & 100  & - & - & 4.5e6 \\

        \hline
    \end{tabular}
    \caption{Parameters used for modeling the spectral density of the EM fluctuations.}
    \label{table1}
\end{table}

We first fit the EM fluctuation spectra in the absence of CoFeB, as shown in Fig. \ref{figure3}(a). The theoretical model is shown as a continuous bold line over the experimental data, and the corresponding modeling parameters are listed in Table \ref{table1}. These modeling parameters indicate that, without the CoFeB layer, the dominant sources of fluctuations are the intrinsic nuclear spins and divacancy centers. EM fluctuations arising from superparamagnetic domain flipping $S_{dom}(f)$ and other sources $S_{other}(f)$, appear after the deposition of the CoFeB layer. Fig. \ref{figure3}(a) shows the theoretical modeling of EM fluctuation spectra for the CoFeB deposited diamond sample at 295 K and 125 K. From the modeling parameters in Table \ref{table1}, it is evident that the superparamagnetic domain flipping rate $1/\tau_d$ decreases as the temperature is lowered. The interaction strength $\Delta_{SP}$ increases correspondingly due to enhanced low frequency EM fluctuations caused by the slower domain flipping rate. In contrast to the temperature dependence, varying the magnetic field produces changes only in the correlation time of the EM fluctuations associated with the nuclear spins and divacancy centers ($\tau_c$) as shown in Fig. \ref{figure3}(b). The peak at 1.5 MHz arises from spin echo modulation due to the nitrogen nuclear spin. A comparison of the EM fluctuations spectra for ferromagnetic and superparamagnetic CoFeB at 295K and 125K is shown in Fig. \ref{figure3}(c). Unlike superparamagnetic CoFeB, ferromagnetic CoFeB shows no EM fluctuations originating from magnetic domain flipping. The spectral analysis demonstrates the unique behavior of superparamagnetic EM fluctuations relative to other types of sources.

\subsection{Distance dependence of the near-field EM fluctuations in CoFeB}
\begin{figure}[h!]
\centering\includegraphics[scale = 0.64]{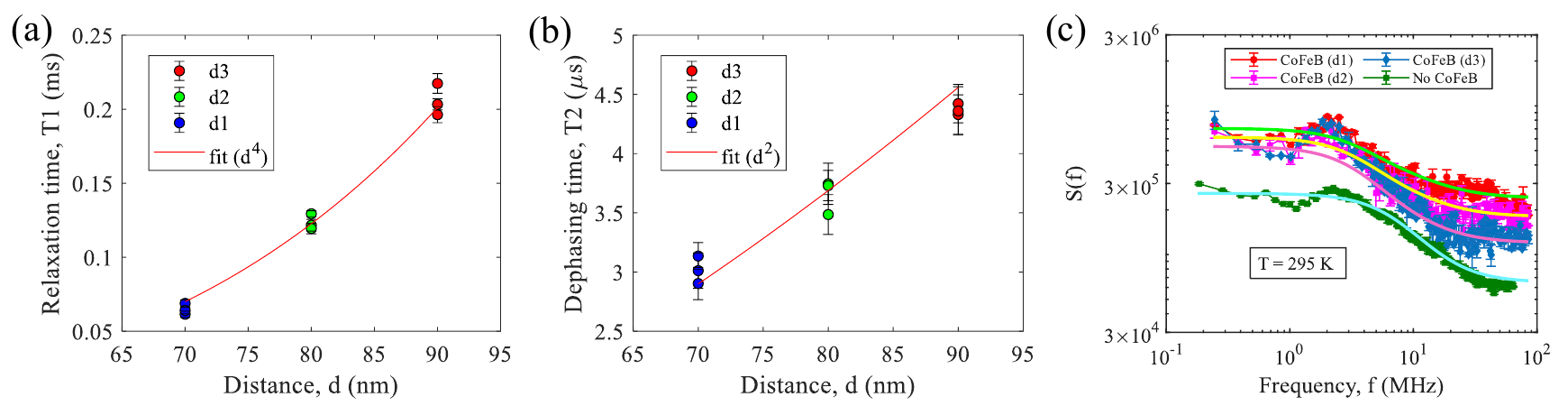}
\caption{Dependence of the coherence times of NV centers on NV-to-CoFeB distance. (a) Relaxation time of NV centers as a function of their distance from the CoFeB layer. The relaxation time follows a distance-dependent scaling of $d^4$. (b) Dephasing time as a function of distance from the CoFeB layer. The dephasing time exhibits a $d^2$ dependence on the distance. (c) EM fluctuations spectrum measured at various distances from the CoFeB layer. The spectrum reveals how the amplitude at different frequency components varies with distance from the magnetic source.}
  \label{figure5}
\end{figure}

We investigate the distance dependence of superparamagnetic near-field fluctuations to understand how the distance and geometry of the magnetic layer affect the relaxation and dephasing dynamics of NV spins. Fig. \ref{figure5}(a) presents the longitudinal relaxation time $T_1$ of the NV centers as a function of the distance between the CoFeB layer and the NV layer, for the device shown in Fig. \ref{figure5}(a). The data clearly reveal a scaling of $d^4$ of $T_1$ with the separation distance $d$. To verify the spatial uniformity of the relaxation time in the diamond sample, measurements were made at each distance at three different locations. This scaling behavior can be explained by the geometry of the CoFeB layer. The magnetic moments of spins in the CoFeB layer are dipolarly coupled to the NV centers. The fluctuating magnetic field from a single spin scales as $\langle B^2(\tau)\rangle \propto \frac{1}{r^6}$ \cite{pham2016nmr}, where $r$ is the distance between the magnetic spin and the NV center. Since the effective thickness of our CoFeB layer, $t_{\mathrm{CoFeB}} = 1.1\,\mathrm{nm}$, is much smaller than $d$, we can treat the layer as a quasi-two-dimensional structure. In this case, the total magnetic field variance from the layer can be expressed as an area integral - 
\begin{equation}
\langle B_{\mathrm{rms}}^2 \rangle = \int \langle B^2\rangle\, dA \propto \frac{1}{d^4}.
\end{equation}
This theoretical scaling is consistent with our experimental observations.

Next, we measured the dependence of the dephasing time $T_2$ on the CoFeB-NV separation d. The results, presented in Fig. \ref{figure5}(b), demonstrate a scaling behavior of $d^2$. This scaling is slower than the expected dependence on $d^4$. The reason for this discrepancy is the contribution of intrinsic nuclear spin noise, which is independent of the distance from the CoFeB layer. We subtracted the intrinsic noise, obtained from measurements without the CoFeB layer, from the total noise observed in the presence of the layer. However, there is a residual effect of the nuclear spins on the observed dephasing time. To further explore how distance influences specific noise components, we performed a spectroscopic decomposition of the magnetic noise spectrum at various separations (Fig. \ref{figure5}(c)). The theoretical models for the spectra (using equation \ref{Noise_All}) are shown on the experimental data as a continuous bold line.  The corresponding model parameters are shown in Table \ref{table2}. The analysis shows that the nuclear spin background remains approximately constant, whereas noise contributions from domain wall motion and superparamagnetic spin fluctuations decrease with increasing distance. These components gradually follow a faster scaling trend with higher power than $d^2$ at higher frequencies. This is attributed to the reduced influence of nuclear spin noise at higher spectral frequencies. These studies reveal the spatial distribution of the fluctuating field due to a superparamagnetic nanoscale magnetic layer, which can be important for spintronics device applications.

\begin{table}[h!]
    \centering
    \begin{tabular}{c|c|c|c|c|c|c|c}
        Sample &  T(K) & d(nm) & D (MHz) & $\tau_c$(ns) & $\Delta_{SP}$ (MHz) & $\tau_d$(ns) & $S_{other}$\\
        \hline \hline
          & 295 & 70 & 5.5 & 40  & 5   & 10  & 2.4e5 \\
        Superparamagnetic CoFeB  & 295 & 80 & 5.5 & 40  &  4 & 10  & 1.8e5 \\
          & 295 & 90 & 5.5 & 40  &  3 &  10 & 1.2e5 \\
        \hline   
    \end{tabular}
    \caption{Parameters used for modeling the distance dependence of superparamagnetic EM spectral density.}
    \label{table2}
\end{table}

\section{Conclusion}

This work presents the first demonstration of the quantum sensing of superparamagnetic behavior in a nanoscale CoFeB layer within an integrated structure. Conventional methods for magnetization characterization are often invasive or impractical in such integrated platforms. In contrast, our approach enables non-perturbative, non-invasive probing of superparamagnetic dynamics. We report observation of unconventional scaling in low-frequency magnetic fluctuations in the CoFeB film, which is corroborated by theoretical analysis and modeling. Through noise spectroscopy and systematic measurements of magnetic field and distance dependency, we gain key insights into superparamagnetic spin dynamics relevant for hybrid quantum devices. This work constitutes a significant advancement in NV-center-based quantum noise spectroscopy, offering a novel route toward on-chip quantum probes for the characterization of nanoscale magnetic layers. It also shows the pathway for integration of quantum probes into spin-based hybrid quantum technologies.

\section{Experimental Section}
\threesubsection{NV confocal noise spectroscopy measurements}\\
We have conducted our measurements in a home-built confocal microscopy setup. The excitation source for the initialization and readout of NV centers is a 532 nm beam generated by a continuous-wave (CW) solid-state laser. For pulsed measurement, this beam is modulated by an acousto-optic modulator (AOM) controlled via a programmable TTL pulse generator. The laser is focused onto the NV center sample using an objective lens with a numerical aperture (NA) of 0.6. The objective lens has a maximum working distance of 4 mm and is collar-corrected for the 1 mm thick glass window of the cryostat. The lens focuses the green laser to a spot size of approximately 500 nm on the diamond sample. The diamond sample is mounted on the cold finger of the cryostat using a printed circuit board (PCB). A copper antenna on the PCB supplies an RF signal to the NV center sample. The RF signal is generated from a Keysight M8190A arbitrary waveform generator. The signal is amplified using a ZHL-16W-43-S + high-power amplifier. Variable magnetic fields are applied to the NV center sample by placing appropriate magnets around the cryostat snout. Photoluminescence (PL) signals emitted from the diamond sample are collected with the same objective lens and pass through a dichroic mirror with a cutoff wavelength of 550 nm. The PL signal is detected using a pair of Micro Photon Devices SPADs. Higher spatial resolution is obtained using a pair of lenses with a pinhole aperture to collect only fluorescence from the objective's focal plane. The time-tagged PL signal is processed by a Hydraharp time-correlated single-photon counting (TCSPC) device. This time-tagged signal is used to read the spin states of NV centers during various measurements and to create confocal images.\\\\
\threesubsection{Diamond-CoFeB sample preparation and characterization}\\
In order to create a near-field interaction platform between spintronic materials and NV centers in diamond, we directly deposited an ultrathin layer of amorphous CoFeB on the diamond surface. The diamond substrate used in this work is an optical-grade chemical vapor deposition-grown (CVD grown) diamond slab commercially available from Element Six with dimensions $4\times4\times0.5$ mm. We first etched $5\ \mu m$ on each surface of the diamond substrate in an inductively coupled plasma etcher using $Ar$, $O_2$, and $Cl_2$ to release the surface stress formed during CVD growth and polishing. Diamond surface roughness was reduced to $\sim 500\ pm$ rms post-etching. The diamond substrate is then cleaned in Piranha solution to remove organic residue and graphitized diamond, followed by standard solvent cleaning to remove inorganic contamination. In order to form the NV layers, the diamond sample is implanted with nitrogen ions (done by CuttingEdge Ions) with a concentration of $10^{12}\ ions\ cm^{-2}$. 

In order to preserve the NV properties and to control the NV-material distance, we sputtered silica spacer directly on the surface of our diamond substrate. We deposited SiO$_{2}$ with thicknesses of 20, 30, 40 nm on the same substrate using a shadow mask placed on the diamond (details in supplementary material section 2). The magnetic material was deposited onto the silica spacer on diamond using a stoichiometric target Co$_{60}$Fe$_{20}$B$_{20}$ in a physical vapor deposition (PVD) sputtering chamber with a base pressure of $10^{-6}$ Torr for magnetron sputtering. The CoFeB was deposited with a 30 W power in 15 sccm of Ar gas for 4 minutes, and was immediately encapsulated by 15 nm of Alumina sputtered in the same chamber. The magnetization measurements of the CoFeB-deposited diamond was performed in the Quantum Design MPMS-3 EverCool SQUID magnetometer using vibrating sample magnetometry (VSM). The VSM measurements was conducted both along the in-plane and normal direction to the diamond surface to characterize the magnetization of the as-deposited CoFeB in an applied magnetic field.


\medskip
\textbf{Supporting Information} \par 
Supporting Information is available from the Wiley Online Library or from the author.

\medskip
\textbf{Acknowledgements} \par 
This work is supported by the Army Research Office (W911NF-21-1-0287).

\medskip

%

\printbibliography

\end{document}